\documentclass{article}

\usepackage{arxiv}

\usepackage[utf8]{inputenc} 
\usepackage[T1]{fontenc}    
\usepackage{hyperref}       
\usepackage{url}            
\usepackage{booktabs}       
\usepackage{amsfonts}       
\usepackage{nicefrac}       
\usepackage{microtype}      
\usepackage{lipsum}
\usepackage{amsmath}
\usepackage{array}
 \usepackage[table]{xcolor}
\newcolumntype{P}[1]{>{\centering\arraybackslash}p{#1}}

\title{Towards Decentralization of Social Media}

\author{
  Sarang Mahajan \\
  Department of Computer Engineering\\
  Pune Institute of Computer Technology\\
  Pune, India \\
  \texttt{smsmahajan34@gmail.com} \\
   \And
 Amey Kasar \\
  Department of Electronics and Telecommunication Engineering\\
  Pune Institute of Computer Technology\\
  Pune, India \\
  \texttt{kasaramey16@gmail.com} \\
}

\begin{document}

\maketitle

\begin{abstract}
 Facebook uses Artificial Intelligence for targeting users with advertisements based on the events in which they engage like sharing, liking, making comments, posts by a friend, a group creation, etcetera.  Each user interacts with these events in different ways, thus receiving different recommendations curated by facebook’s intelligent systems.  Facebook segregates its users into chambers, fragmenting them into communities[4]. The technology has completely changed the marketing domain. It is however caught in a race for our finite attention with a motive to make more and more money. Facebook is not a neutral product. It is programmed to get users addicted to it with a goal of gaining added information about the users and optimizing the recommendations provided to the users according to his or her preferences. This paper delineates how Facebook’s recommendation system works and presents three methods to safeguard human vulnerabilities exploited by Facebook and other corporations. 

\bigbreak

The primary focus of the proposed work is to describe a social media platform called Anti-Facebook to safeguard users' privacy and data using the concept of decentralization. In the past, data collected from Facebook has been used to influence global events significantly. The motivation of this paper is to prevent the use of the data garnered from Facebook to alter the beliefs held by people by influencing their minds. 

\bigbreak
The rapid growth of social networks has led to burgeoning infrastructure that has required platform owners to primarily earn their keep through ad sales. This has led users to be at a disadvantage because their data has become the “currency” being traded by social networks. A decentralized approach to social networking can ensure better privacy, as well as the potential to do more with smart apps and contracts, as well as e-commerce and crowdfunding transactions.
\end{abstract}


\section{Introduction}
Advertisements optimized by Artificial Intelligence are different than regular advertisements we experience because the optimized advertisements are not based on a singular relationship but a two way relationship[1]. In this scenario of optimized advertisements, they are not just random advertisements shown to any user but they are subtly manipulated by using Deep Neural Networks to approximate the content that is most likely to be relevant according to user’s preferences and the same technique is used to display advertisements to users in order to make users click them[3]. Facebook provides free services to its users hence, generating revenue through advertisements is a necessity for them. To improve their profits and supply the most relevant advertisements to its users to capture their attention, Facebook needs to know the preferences of its users which it determines by analysing the users data. To capture our attention the system needs to us to engage with us from time to time which is not good for our well being. The race to keep us engaged makes it harder for us to disconnect, increasing stress, anxiety and reducing sleep. Facebook constant usage trains children to replace their self-worth with the number of likes they get on their posts, encourage comparison with others, and creates a constant illusion of missing out. The race for attention is degrades social relationships among people by  forces them to prefer virtual interactions and rewards (likes, shares on their posts) over face-to-face community. Outrage and false facts are better at gaining attention of people and divides us so we can no longer agree on truth. With millions of active users on facebook, societies are easily manipulated with unprecedented precision. 
\bigbreak

Facebook makes it easier than ever for bad actors to cause havoc. Using Facebook’s advertising system bad information can be directly pushed to users with specific zip codes, races, or religions. This makes it easy to target people who are already prone to certain conspiracies or social issues. The content delivered is timed to prey on the users when they are most vulnerable. Millions of fake accounts and bots impersonating real people with real sounding names and photos are used to fool millions by generating a false sense of consensus. Because all of this, the platform companies profit from growth in users and activities. We cannot expect attention extracting company like facebook to change because it is against their business model and would lose money if they blocked advertisers from micro-targeting lies and conspiracies to the people most likely to be persuaded. 
\bigbreak
		\setlength{\tabcolsep}{20pt}
	    \setlength{\arrayrulewidth}{0.35mm}

		\begin{table}[h!]
		  \centering
	    {\rowcolors{2}{green!20}{green!15}
		\begin{tabular}{ |P{1.35cm}|P{1.35cm}|P{1.35cm}|P{0.5cm}|P{1.35cm}|P{1.35cm}|  }
		
		\hline
    
	Users  & Event 1 & Event 2 & .... & Event l-1 & Event l  \\
	\hline
	User 1           & 0 & 0 &  ....  & 4 & 2      \\ 
	User 2           & 0 & 0 &   .... & 4 & 2     \\ 
	User 3           & 0 & 0 &   .... & 4 & 2  \\ 
	User  4          & 0 & 0 &  ....  & 4 & 2  \\ 
	User 5           & 0 & 0 &  ....  & 4 & 2       \\
	User 6           & 0 & 0 &   .... & 4 & 2 \\
    ........         &   &   &   .... &   &                \\
	User n-2         & 0 & 0 &  ....  & 4 & 2       \\                       
	User n-1         & 0 & 0 &  ....  & 4 & 2 \\
	User n           & 0 & 0 &  ....  & 4 & 2 \\
	\hline
	\end{tabular}}
	  \newline\newline
  \caption{title}\label{tab1}
	\end{table}
	
	\bigbreak
We can represent all the events users engaged themselves into as numerical values and represent it as a rating matrix as shown in fig. The rows are all the unique users and columns are all the possible events. The assumption in Facebook’s recommendation system is that users who like, share similar posts or view the same articles share one or more hidden preferences and are likely to react in a similar fashion to same content. Thus, Facebook’s recommendation system relies solely on the user behavior to make recommendations which is known as collaborative filtering.  Every entry in the Facebook’s data matrix represents a user’s reaction to events. If a user has never engaged in a particular event then the rating for that event will be zero. The data matrix is sparse in nature and thus, most of the entries in it are zero. Each user and events are latent features, such that dot products of these vectors closely match known preferences. By feeding this data to a Neural Network, the preferences of unknown users can be approximated. The objective function in this case that we want to minimize is :
\bigbreak
\[
min \sum_{ratings\hspace{0.5em} u,i} (r_{u,i}  - x_u \cdot y_i)^2 + \gamma  \bigg(\sum_{users\hspace{0.5em} u} \lvert x_{u} \rvert^2 +\sum_{items \hspace{0.5em} i} \lvert y_{i} \rvert^2 \bigg) 
\]

\bigbreak
Here, r are the known user ratings, and x and y are the user and item feature vectors that we are trying to find. As there are many free parameters in the equation 1, we need the regularization part to prevent overfitting, with gamma being the regularization factor. Training the machine to learn the pattern of user preferences is carried out by iterative approach that starts by generating random weights that are used to dot product with rating matrix after summing its latent features and gradually improve the solution. Stochastic gradient descent (SGD) optimization is used to loop through the ratings and make prediction for each rating and compute the error in prediction with an aim to reduce the error in next iterations by updating the weights. The problem with this approach is that millions of ratings and users which makes it very difficult to process that data for each iteration during training.
\bigbreak
The method to overcome the data harvesting process of Facebook is put forth in section 2. The outcomes expected from proposed theory are delineated in section 3. Section 4 presents the improvements than can be made in future pertinent  to this paper.

\section{Proposed Theory}
\label{sec:headings}

The short term solution to avoid being manipulated by facebook is to avoid social media. It can be done by turning off notifications except from people. Notifications appear in red color because red is a trigger color that instantly draws attention. But most notifications are generated by machines, not actual people. They keep making alert sounds and keep our phones vibrating to lure us back into apps we don’t really need. Colorful awards give our brains shiny rewards every time we unlock our phone. Setting the phone color to grayscale removes those positive reinforcements and helps people check their phone less. Keeping the first page of apps to tools that we need for in-and-out taks is also a solution to avoid insignificant contact with the phone. Swiping down and typing the the name of the application we want to open instead of scrolling takes enough effort to make us pause and ask ourselves “do I really want to open this app?”
\bigbreak
The second solution to the problem is using a decentralized social media platform like peepeth which encourages mindful and responsible engagement[5]. It is built on top of a Blockchain architecture the gives its users ownership control. It encourages thoughtful content while discouraging reactive and hateful posts. It gives us control of our online legacy as data is saved to Ethereum blockchain which is open, decentralized, and immutable. Peepeth helps to reduce invasion of our privacy and selling of our data. 
\bigbreak
The other solution to this is that Facebook could leverage itself into a decentralized platform. It will enable users to have crypto addresses not owned by anyone. Facebook can also lunch its own crypto currency that can be used to reward their users in exchange of their data. It will lead to surge of Facebook because its new business model will prevent users from viewing optimized preferences and will only show the content that users wish to view.

\section{Expected Outcomes}
\label{sec:headings}

If the version of Facebook which feeds on data of its users for recommending content optimized by Artificial Intelligence stops then people are likely to feel less addicted to Facebook. If Facebook gets decentralized as mentioned in the proposed theory then users will look at Facebook with a view of generating money on small basis which will keep facebook active but on a user friendly basis.
\bigbreak
Facebook simply has too much control over the data of its users, giving it too much power to dictate terms. Having a decentralized infrastructure of blockchain based system provides us with a more balance of power between the platform and its users. Facebook is an easy target for those who want to misuse data due to its centralized data stores, otherwise misappropriating data is far more difficult without a central point of weakness attack.
\bigbreak
Most compelling justification for a decentralized social network is user control of personal data. Having granular control over the data and information shared with applications and websites makes our data and privacy more secure. The additional possibilities with user-based data ownership, such as the ability to monetize it for personal gain, but it’s a completely different dynamic when it’s a personal choice, not a corporate business model.
\bigbreak
Another important component of a decentralized social networking ecosystem is that of portability. The idea is to provide users with the same type of functionality without the same degree of one-sided data control. This means that a user’s personal identity remains in their control and can be moved from one app to another or exchanged freely between different platforms. The user’s profile exists separately from the application or service and interfaces with it only as much as is necessary. The onus is on the user to protect their individual data, but there is no longer a concern that a third-party intermediary will be able to sell, harvest or misuse it.
\bigbreak
If aggregated analytics and user targeting are not bad enough, social networks like Facebook are even being accused of eavesdropping on users through smartphone microphones.For most people, this might not be a concern, but for those who are paranoid about their privacy, then the most viable option is to not join social networks at all. However, social networks still have their merits, in terms of business networking, collaboration, and exchange of ideas.
\bigbreak
For those in repressive regimes or where censorship is an issue, a blockchain-based approach to social networking offers the benefits of secure authentication whilst still ensuring anonymity. Even with messaging services like iMessage, WhatsApp, and others, having end-to-end encryption, the problem lies with the meta-data that gets exchanged with the messages, which leave digital breadcrumbs that third parties can pick up. Thus, even if eavesdroppers do not know the contents of a message, they can determine where it came from, who it is addressed to, and other such details.

\section{Future Scope}
\label{sec:headings}
There has not been a real implementation of a social media platform apart from peepeth that uses blockchain to store data of its users in order to protect privacy and data. In future, we look forward to address this issue by spreading more awareness about harm caused by centralized version of Facebook and look to come up with a decentralized version of social media.


\begin{thebibliography}s

\bibitem{c1}Explaining the News Feed Algorithm : An Analysis of the "News Feed FYI" Blog, Kelley Cotter, Janghee Cho and Emilee Rader.

\bibitem{c2}The Anti Facebook,Siraj Raval, https://www.youtube.com/watch?v=T5zIlWSMlU8

\bibitem{c3}Recommender Systems Explained, Pavel Kordik, https://medium.com/recombee-blog/recommender-systems-explained-d98e8221f468

\bibitem{c4} Our Society is being hijacked by technology, https://medium.com/recombee-blog/recommender-systems-explained-d98e8221f468

\bibitem{c5} A new kind of social network, https://peepeth.com/welcome.




\end{thebibliography}
\end{document}